\documentclass[reprint,aps,prd,twocolumn,showpacs,showkeys, 10pt, longbibliography, nolinenumbers, floatfix]{revtex4-1}

\usepackage{amsmath}
\usepackage{amssymb}
\usepackage{amsfonts}
\usepackage{graphicx}
\usepackage{float}
\usepackage{dcolumn}
\usepackage{multirow}
\usepackage{natbib}
\usepackage{tcolorbox}
\usepackage{parskip}
\usepackage[colorlinks = true,linkcolor = blue,urlcolor  = blue,citecolor = blue,anchorcolor = blue]{hyperref}
\usepackage{enumerate}

\usepackage{times}
\begin{document}

%\usepackage{stfloats} %allow table* bottom placement 
%\usepackage{newtxmath}
%usepackage{upgreek}

\title{Quarkyonic equation of state with  momentum-dependent interaction  and neutron star structure}
\author{K. Folias and Ch.C. Moustakidis}
%\email{alkanaki@auth.gr}
%\author{Veronica Dexheimer$^2$}

%\author{C.C. Moustakidis$^1$}
%\email{pkoliogi@physics.auth.gr}

%\author{C.C.C}
%\email{moustaki@auth.gr}

\affiliation{Department of Theoretical Physics, Aristotle University of Thessaloniki, 54124 Thessaloniki, Greece 
%$^2$Department of Physics, Kent State University, %Kent, OH 44243 USA }
}
%%%%%%%%%%%%%%%%%%%%%%%%%%%%%%%%%%%%%%%%%%%%%%%%%%%%%%%%%%%%%%%%%%%%%%%%%%%%%%%%
\begin{abstract}
%%%%%%%%%%%%%%%%%%%%%%%%%%%%%%%%%%%%%%%%%%%%%%%%%%%%%%%%%%%%%%%%%%%%%%%%%%%%%%%%	
The structure and basic properties of dense nuclear matter still remain one of the open problems of Physics. In particular, the composition of the matter that composes neutron stars is under theoretical and experimental investigation. Among the theories that have been proposed, apart from the classical one where the composition is dominated by hadrons, the existence or coexistence of free quark matter is a dominant guess. An approach towards this solution is the phenomenological view according to which the existence of quarkyonic  matter plays a dominant role in the construction of the equation of state (EOS). According to it the structure of the EOS is based on the existence of the quarkyonic particle  which is a hybrid state of a particle that combines properties of hadronic and quark matter with a corresponding representation in momentum space. In this paper we propose a phenomenological model for quarkyonic matter, borrowed from corresponding applications in hadronic models, where the interaction in the quarkyonic  matter  depends not only on the position but also on the momentum of the quarkyonic particles. This consideration, as we demonstrate, can have a remarkable  consequence on the shape of the EOS and thus on the properties of neutron stars, offering a sufficiently flexible model.  Comparison with recent observational data can place constraints on the parameterization of the particular model and help improve its reliability.

%\pacs{26.60.-c, 26.60.Kp, 97.60.Jd}

\end{abstract}

\maketitle

%%%%%%%%%%%%%%%%%%%%%%%%%%%%%%%%%%%%%%%%%%%%%%%%%%%%%%%%%%%%%%%%%%%%%%%%%%%%%%%%
%\end{abstract}
%%%%%%%%%%%%%%%%%%%%%%%%%%
%\section{Introduction}
%%%%%%%%%%%%%%%%%%%%%%%%%%%

%%%%%%%%%%%%%%%%%%%%%%%%%%%%%%%%%%%%%%%%%%%%%%%%%%%%%%%%%%%%%%
\section{Introduction}
%%%%%%%%%%%%%%%%%%%%%%%%%%%%%%%%%%%%%%%%%%%%%%%%%%%%%%%%%%%%%%
One of the fundamental problems of Physics remains the composition of dense nuclear matter as well as its basic properties both at zero and at finite temperature~\cite{Glendenning1996CompactSN,Shapiro:1983du,Haensel2007NeutronS1,schaffner-bielich_2020,Moustakidis2007ThermalEO,PhysRevC.91.035804,2023PhRvD.107d3005L,PhysRevD.107.083023,Ming-2023,Fan-2024,Bombaci_2018,2020GReGr..52..108B,Metzger:2019zeh,2020LRR....23....4B,Piekarewicz:2013dka,Vi_as_2021,Oikonomou_2021,2024Symm...16..658M}. In particular, the  equation of state   of neutron star matter is the key quantity to study  these objects. In recent years very important astrophysical observations concerning not only isolated neutron stars but also binary systems, such as for example the merger of such a system with parallel emission of gravitational waves, can help decisively in the deeper knowledge of the interior of these objects. In this effort, a key problem that often arises is the inability of the EOSs to predict maximum masses for neutron stars that are compatible with recent observations (well above two solar masses) without simultaneously violating the sound-speed causality.

An interesting attempt in this direction is the consideration of a hybrid state of dense nuclear matter called quarkyonic matter~\cite{PhysRevC.110.025201,PhysRevC.107.065201,Cao:2022inx,Kovensky:2020xif,Quiros:1999jp,Das:2000ft,Koch_2023,Jeong:2019lhv,PhysRevD.107.074027,PhysRevLett.122.122701,McLerran:2020rnw,article,Fukushima_2016,PhysRevC.108.045202,PhysRevLett.132.112701,McLerran-2024,Kojo-2024,Bluhm-2024}. In this consideration,  quarkyonic matter  both quarks and nucleons appear as quasiparticles. Following the analysis of Refs.~\cite{PhysRevLett.122.122701,McLerran:2020rnw,article} the basic assumption of quarkyonic matter is that at large Fermi energy, the degrees of freedom inside the Fermi sea may be treated as quarks, and confining forces remain important only near the Fermi surface where nucleons emerge through correlations between quarks. In this case one can consider that quarks, confinement at the Fermi surface,  occupying a momentum shell of width $\Delta \simeq \Lambda_{QCD}$ (where $\Lambda_{QCD}$ represents  the confinement
scale),   produce triplets with spin $1/2$, that are the baryons \cite{PhysRevLett.122.122701,Fukushima_2016,PhysRevC.108.045202,PhysRevLett.132.112701,McLerran-2024}. The width of the momentum shell $\Delta$ depends on the baryon density.

The motivation of the present work is to propose a quarkynonic model  where the interaction between  baryons, depends not only on the position but also on the momentum.
In this case, the contribution on the energy density and pressure of the potential energy it depends, as in the case of kinetic energy, on the distribution of baryons and quarks in the momentum space. This approach enhances and further completes the quarkyonic model.
The proposed  assumption leads to a model that is even more flexible in describing the properties of dense nuclear matter.
In particular, we consider that in the quarkyonic matter (QM), in the regions near the Fermi surface, there is an interaction between the baryons, which is of the Skyrme type. Such an interaction would likely impact the properties of QM and seemingly influence the overall equation of state as well. 
This approach allows to adjust the stiffness of the equation of state without violating causality in the speed of sound. Another benefit of this model is that it can be extended from zero to finite temperature. In this scenario, the thermal effect will influence not only the kinetic energy term but also the interaction term. 
As far as we know, such an extension of QM has not been attempted before. Nonetheless, we believe that extending and applying the proposed model could provide valuable insights.

In the current study, we examine a QM scenario that includes neutrons alongside up and down quarks. We account for the interaction of neutrons as being dependent on their momentum, while quarks are treated as non-interacting and follow the statistics of a free Fermi gas. We develop a set of parameterized equations of state for quarkyonic matter, tailored to describe neutron star matter.
The parameterization focuses on certain microscopic phenomenological parameters of classical QM. The impact of these parameters on the properties of neutron stars is thoroughly presented and analyzed. Particular emphasis is placed on correlating the QM parameters with the tidal deformability of neutron stars, a critical factor for deducing information about the interior structure of neutron stars. This quantity is especially important because it can now be measured observationally through the analysis of gravitational waves detected from neutron star binary mergers.
One of the primary aims of this work is not only to reproduce the fundamental properties of neutron stars using this model but also to determine whether possible constraints can be placed on its parameterization. Specifically, efforts were made to introduce certain limitations on the model's parameterization by utilizing astrophysical observations. 

The paper is organized as follows: In Section II, we introduce two simple theoretical models, a pure neutron matter model and the  NDU (neutrons, up and down
quarks)  quarkyonic model, both with a momentum-dependent interaction. In Section III we present a short introduction on neutron star structure. In Section IV, we discuss the results of the study, and in Section V, we conclude with our final remarks.

\section{The model}
\begin{figure}
\includegraphics[width=180pt,height=15.0pc]{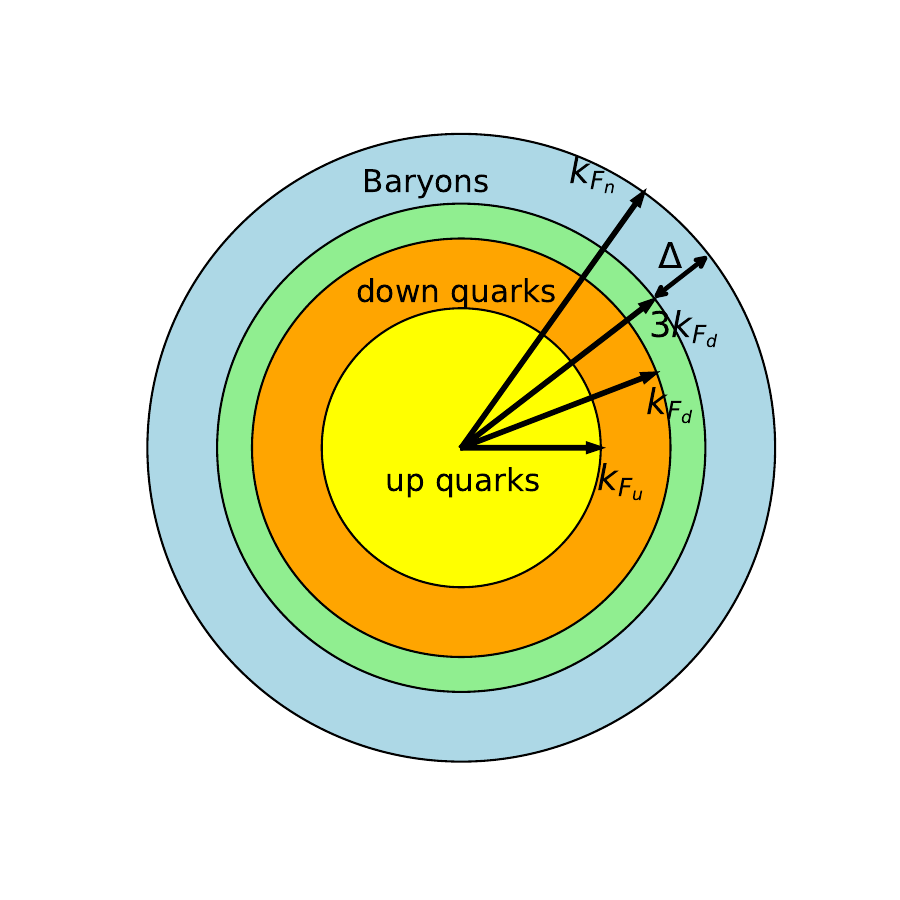}
\caption{The momentum space of quarkyonic matter. Quarks occupy all states from zero momentum up to quark Fermi momentum ($k_{F_d}$ and $k_{F_u}$ for the down and up
quark respectively). At higher momentum, quarks are confined inside baryons that occupy momentum states up to fermi momentum $k_{F_n}$ in a shell with inner radius
$3k_{F_d}$ and with width $\Delta=k_{F_n}-3k_{F_d}$ (see also the text and Ref.~\cite{PhysRevLett.122.122701} for more details).}
\label{fig1}

\end{figure}

%%%%%%%%%%%%%%%%%%%%%%%%%%%%
In order to have a useful overview of the QM model we have included Fig.~\ref{fig1}. This structure, where the low and high momentum states occupied by quarks and neutrons respectively, will help below to better understand the contribution of the QM components (neutrons and quarks) to the total energy density. We will now proceed by presenting two cases: the case of pure neutron  matter and the case of NDU matter with momentum-dependent interaction. 

\subsection{Pure neutron matter}
%%%%%%%%%%%%%%%%%%%%%%%%%%%%%%%%
We begin our calculations with a pure neutron model  at zero temperature to establish a comparison with the quarkyonic model. First, we need to compute the number density of neutrons, which will be in the form 
\begin{equation} n_n = g_s\int_0^{k_{F_{n}}} \frac{d^3k}{(2\pi)^3} =\frac{g_s}{6\pi^2}k_{F_{n}}^3 
\label{eq1}
\end{equation}
%%%%%%%%%%%%%%%%%%%%%%%%%%%%%%%%%%%
where $g_s$ is the degeneracy of the spin and is equal to $2$. In the present study we will use the Fermi momentum $k_{F_n}$ of the neutrons as a key variable in the calculations, instead of the neutron number density $n_n$.
Now,  the energy density of neutrons is given by 
\begin{equation}{\cal E}_{n}(k_{F_n}) =\frac{g_s}{2\pi^2}\int_0^{k_{F_n}} k^2\sqrt{(\hbar c k)^2+ m_{n}^2c^4}dk + V_{\rm int}(k_{F_n})
\label{eq2}
\end{equation}
where $V_{\rm int}(k_{F_n})$ is the potential energy.
In the present study, in addition to what has been done so far, we will consider that  the potential term $V_{\rm int}(k_{F_n})$ depends not only on the density of neutrons but also on their momentum. In particular we use a  model  which has bee extensively applied in the literature and has the form (see Refs.~\cite{Prakash-1997,PhysRevC.79.045806} and references therein)
 \begin{eqnarray}
V_{\rm int}(k_{F_n})&=&\frac{1}{3}An_0\left(1+x_0   \right)u^2 +\frac{\frac{2}{3}Bn_0\left(1-x_3\right)u^{\sigma+1} }{1+\frac{2}{3}B'n_0\left(1-x_3   \right)u^{\sigma-1}}\nonumber\\
&+& u\sum_{i=1,2}\frac{1}{5}\left[\frac{}{}6C_i-8Z_i \right]{\cal J}_n^i(k_{F_n})
\label{eq5}
\end{eqnarray}
%%%
where 
\begin{eqnarray}
{\cal J}_{n}^i(k_{F_n})&=&\frac{2}{(2\pi)^3}\int{\rm g}(k,\Lambda_i)  d^3k
\label{Jn-1}
\end{eqnarray}

The primary source of momentum dependence in Brueckner theory arises from the nonlocal nature of the exchange interaction. As discussed by Bertsch et al.~\cite{Bertsch-1988}, a single-particle potential $U(n)$ that depends solely on baryon density is an oversimplification. Moreover, it is well established that nuclear interactions involve significant exchange effects, which introduce a momentum dependence to the single-particle potential, subsequently influencing the energy density functional. To conduct comprehensive studies of heavy ion collisions, Gale et al.~\cite{Gale-1987} proposed the following parametrization for the momentum component of the single-particle potential
\begin{equation}
 U(n,k) \sim C\frac{n}{n_0}\frac{1}{1+(k-\langle k'\rangle)^2/\Lambda^2}   
\end{equation}
%%%%%%
The present model, which is a generalization of that
proposed by  Gale et al.~\cite{Gale-1987}, has been successfully applied
in heavy ion collisions and astrophysical studies over the
years (see Refs.~\cite{Prakash-1997,BAaoAnLi-2021,Li-2008} and references therein).

The first two terms of the right-hand side of Eq.~(\ref{eq5}) arise
from local contact nuclear interactions that led to power density
contributions such as in the standard Skyrme equation of state.
These are assumed to be temperature independent. Moreover,  the first term  represents an attractive interaction, while the second one  corresponds to a repulsive interaction that becomes dominant at high densities (for $n > 0.6$ fm$^{-3}$). 
The third term describes the
effects of finite-range interactions according to the chosen
function ${\rm g}(k,\Lambda_i)$, and is the temperature-dependent part of
the interaction. This interaction is attractive and important at
low momentum, but it weakens and disappears at very high
momentum. The function ${\rm g}(k,\Lambda_i)$, suitably chosen to simulate
finite-range effects and in the present work has the form~\cite{PhysRevC.79.045806}
\begin{equation}
{\rm g}(k,\Lambda_i)=\left[1+\left(\frac{k}{\Lambda_i}  \right)^2 \right]^{-1}  
\label{g-1}
\end{equation}
%%%%%%%%%%%%%%%%%%%%%%%%%%%
By choosing the function above, we introduce two finite-range terms: one representing a long-range attraction and the other a short-range repulsion. (for more details see also Refs.~\cite{Prakash-1997,Bertsch-1988,Gale-1987}). 
Moreover, in Eq.~(\ref{eq5}) we use the notations $u =n_n/n_0$ (where $n_0$ is  the saturation density $n_0 = 0.16$ fm$^{-3} $),  $\Lambda_1 = 1.5k_{F{n_0}}$, $\Lambda_2 = 3k_{F{n_0}}$ (where  $k_{F{n_0}}$ is the neutron Fermi momentum at the saturation density $n_0$). 
The parameterization of  Eq.~(\ref{eq5}) is presented  in Table \ref{tab-1}.
The parameters $A$, $B$, $B'$ $\sigma$, $C_1$ and $C_2$, which
appear in the description of symmetric nuclear matter and the
additional parameters $x_0$, $x_3$, $Z_1$, and $Z_2$ used to determine
the properties of asymmetric nuclear matter, are treated as
parameters constrained by empirical knowledge (for more details see Ref.~\cite{Prakash-1997}). The integral of the expression~(\ref{Jn-1})  can be calculated analytically and gives
\begin{equation}
\int k^2 \left[1+\left(\frac{k}{\Lambda}  \right)^2 \right]^{-1} dk=\Lambda^2\left[k-\Lambda \arctan\left(\frac{k}{\Lambda}\right)  \right]    \label{eq7}
\end{equation} 
%%%%%%%%%%
The above result will help clarify the effect of the momentum-dependent interaction on the quarkyonic equation of state, as discussed below.
\begin{table}
\centering
\caption{The parametrization of the potential energy  $V_{\rm int}(k_{F_n})$ given by Eq.~(\ref{eq5}) (for more details see Ref.~\cite{Prakash-1997}).}
\label{tab-1}    
\begin{tabular}{cccccccccc}
\hline
 $A$  & $B$ & $B'$& $\sigma$ &  $C_1$ & $C_2$ & $x_0$ &  $x_3 $ & $Z_1$ & $Z_2$  \\\hline
-46.65 & 39.45 & 0.3 & 1.663 & -83.84 & 23 & 1.654 & -1.112 & 3.81 & 13.16 \\\hline
\end{tabular}

\end{table}
Furthermore, the chemical potential of neutrons will be determined using the standard thermodynamic relation. \begin{equation} \mu_n = \frac{\partial {\cal E}_n}{\partial n_n}\label{eq3}\end{equation}
and the corresponding  pressure by the expression,
 \begin{equation} P_n= \mu_n n_n - {\cal E}_n\label{eq4}
 \end{equation}
 %%%%%%%%%%%%%%
The energy density ${\cal E}_n$ and pressure $P_n$ determine the EOS of pure neutron matter.  The key quantity in this study is the speed of sound $c_s$ which (in units of the speed of light) is defined as 
\begin{equation}\frac{c_s}{c}=\sqrt{\frac{\partial P_n}{\partial {\cal E}_n}}
\label{speed-1}
\end{equation}

For an equation of state  to respect causality, it must satisfy the following condition:
\[\frac{c_s}{c} \leq 1  \]
in all ranges of densities.

\label{sec II}

%%%%%%%%%%%%%%%%%%%%%%%%%%%%%%%%%%%%
\subsection{The NDU quarkyonic model incorporating momentum-dependent interaction forces}
%%%%%%%%%%%%%%%%%%%%%%%%%%%%%%%%
To investigate quarkyonic matter and its impact on neutron star properties, we begin with a simplified model where the neutron star is composed solely of neutrons, along with up and down quarks~\cite{PhysRevLett.122.122701,Zhao:2020dvu}. This assumption will simplify our calculations and serves as a useful approximation, particularly considering that in a neutron star, the number of protons and electrons is minimal compared to the number of neutrons. As a result, neutrons and quarks play the primary role in determining the structure of a neutron star.

We consider that quarks are non-interacting and only neutrons interact with each other. Also, all of our calculations are at $ T=0 $ (cold equations of state), so we consider a fully degenerated Fermi gas (as the pure neutron case). Moreover, we require that the low momentum states be occupied by quarks, while the higher momentum states are occupied by neutrons (see   a 
schematic plot in Fig.~\ref{fig1}).
For the width of the momentum shell we use the relation (see Ref.~\cite{PhysRevLett.122.122701})
\begin{equation}\Delta = \frac{\Lambda_{Qyc}^3}{\hbar^3 c^3 k_{F_n}^2 }+ \kappa \frac{\Lambda_{Qyc}}{\hbar c N_c^2} \label{eq16}
\end{equation}
%%%%%%%%%%%
and we set  $\Lambda_{Qyc} \approx \Lambda_{QCD}$ and $\kappa =0.3$.
In the present study, we consider a more simple expression for the Fermi momentum of down quarks and neutrons, which is derived by the basic assumption of the quarkyonic scenario~(see Refs.~\cite{PhysRevLett.122.122701,PhysRevC.104.055803})
\begin{equation}
k_{F_d} = \frac{k_{F_n} - \Delta}{3} 
\label{eq15} \end{equation}
In this case, $\Delta$ has the form given by Eq.~(\ref{eq16}). Moreover,   we use the values $\Lambda_{Qyc} = 160, 180, 200$ MeV. In order to ensure charge neutrality we consider that,
\begin{equation}n_d = 2 n_u \label{eq9}\end{equation}
or with respect to Fermi momentum,
\begin{equation}k_{F_d} = 2^{1/3} k_{F_u} \label{eq10}\end{equation}

We calculate the quark number density and energy density using the following expressions
\begin{equation}n_Q = \frac{g_s N_c}{2\pi^2}\sum_{i=u,d}\int_0^{k_{F_i}} k^2 dk \hspace{3mm},\label{eq17}\end{equation}
%%%%%%%%%%%%%%%%%%%%%%%%%%%%%%%%
\begin{equation}{\cal E}_{Q} = \frac{g_s N_c}{2\pi^2}\sum_{i=u,d}\int_0^{k_{F_{i}}} k^2\sqrt{(\hbar c k)^2+ m_{Q}^2c^4}dk \label{eq18}\end{equation}
%%%%%%%%%%%%%%%%%%%%%%%%%
The corresponding expressions for the neutrons are  
\begin{equation}n_n =\frac{g_s}{2\pi^2}\int_{k_{F_n}-\Delta}^{k_{F_n}} k^2 dk \hspace{3mm},\label{eq19}\end{equation}

\begin{equation}{\cal E}_{n} =\frac{g_s}{2\pi^2}\int_{k_{F_{n}}-\Delta}^{k_{F_{n}}} k^2\sqrt{(\hbar c k)^2+ m_{n}^2c^4}dk  + V_{\rm int}(k_{F_n}) \label{eq20}\end{equation}
where the term $V_{\rm int}(k_{F_n})$ is given by Eq. (\ref{eq5}) and the integral ${\cal J}_{n}^i$ is defined as
\begin{eqnarray}
{\cal J}_{n}^i&=&\frac{2}{(2\pi)^3}\int d^3k{\rm g}(n,\Lambda_i)\nonumber\\
&=&\frac{2}{(2\pi)^3}\int_{k_{F_n}-\Delta}^{k_{F_n}} 4\pi \left[1+\left(\frac{k}{\Lambda_i}  \right)^2 \right]^{-1}   k^2 dk  \label{eq26}
\end{eqnarray}
%%%%%%%%%%

\begin{figure}
\includegraphics[width=250pt,height=20pc]{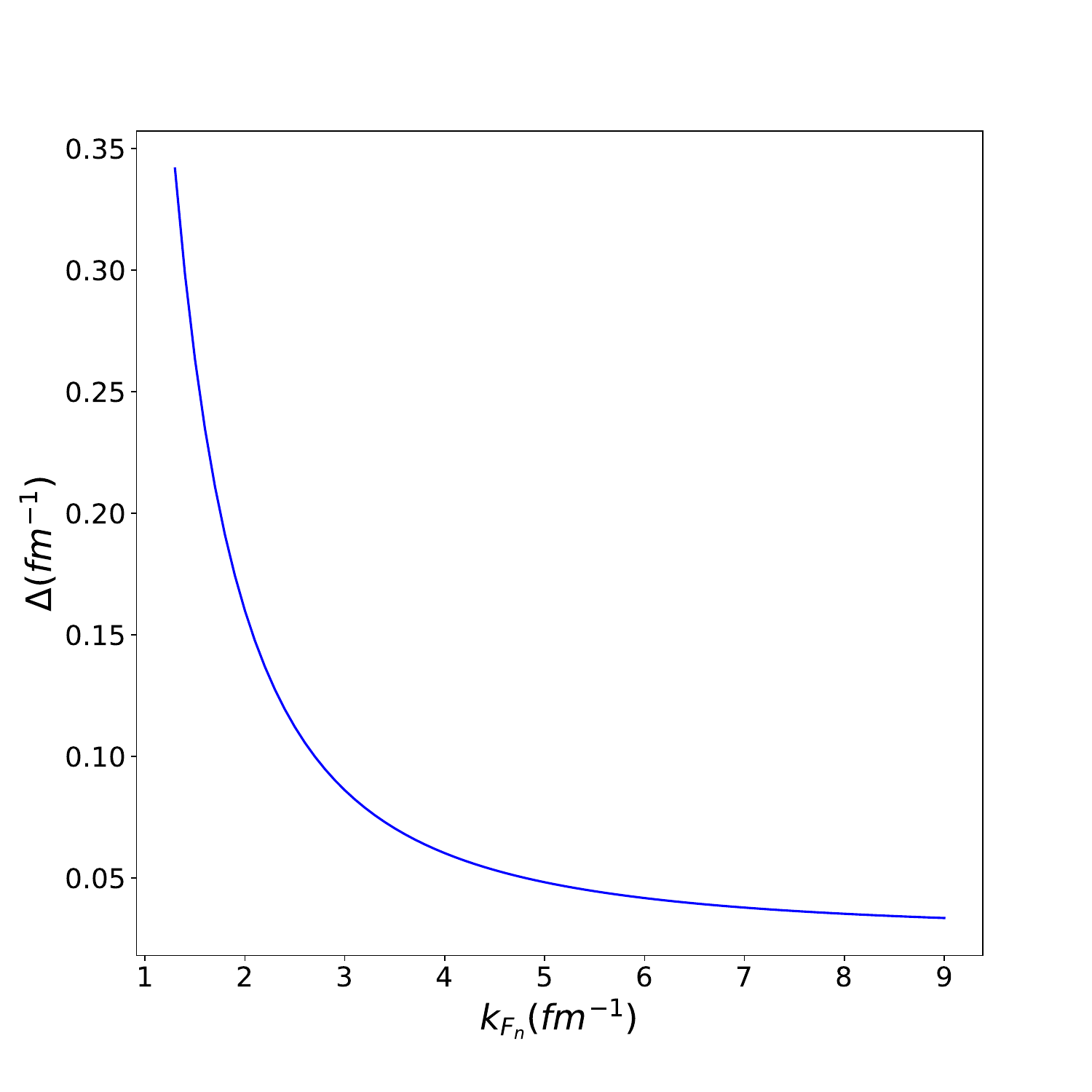}
\caption{The width of the momentum shell ($\Delta$) as a function of neutrons Fermi momentum ($k_{F_n}$), for $\kappa =0.3 $, $N_c = 3$ and $\Lambda_{Qyc} = 160$ MeV .}
\label{Delta}
\end{figure}

We can illustrate the dependence of the width of the momentum shell $\Delta$ as a function of neutrons Fermi momentum $k_{F_n}$ in Fig.~\ref {Delta}. We can easily see that the value of parameter $\Delta$ decreases as the neutron's Fermi momentum and consequently the particle number density increases, which is obvious from Eq.~(\Ref{eq16}) and also it is reasonable if one considers that the greater the Fermi momentum, the greater the number density, so quarks dominate in that region.

The quark chemical potentials will be given by
\begin{equation}\mu_i=\sqrt{(\hbar c k_{F_i})^2+ m_{Q}^2c^4}, \qquad i=u,d 
\label{eq13}
\end{equation}
Relation (\ref{eq13}) arises from the fact that for temperature $T=0$, the chemical potential is equal to the Fermi energy.
Quark masses are obtained by $m_Q = m_N/N_c$ where $m_N$ is the nucleon mass and $N_c$ is the number of colors. We set the number of colors and the degeneracy of the spin equal to $N_c = 3 $ and $g_s = 2$ respectively and we introduce the quarkyonic matter at a baryon density around $n_B = 0.2 - 0.4$ fm$^{-3}$. The total baryon density and total energy density will be respectively

\begin{equation} n_B = n_n + \frac{(n_u + n_d)}{3} = \frac{1}{3\pi^2}(k_{F_{n}}^3- (k_{F_{n}}-\Delta)^3 + k_{F_{u}}^3 + k_{F_d}^3) \label{eq21}\end{equation}

 and

\begin{equation} {\cal E}_{\rm tot}= {\cal E}_n + {\cal E}_Q 
\label{eq22}
\end{equation}

while the chemical potential for each species of matter is defined as 
\begin{equation} \mu_i = \frac{\partial {\cal E}_{\rm tot}}{\partial n_i}, \qquad i=n,u,d 
\label{eq23}
\end{equation}

Now, the total pressure and the speed of sound are  given respectively  by
\begin{equation} 
P_{\rm tot} = - {\cal E}_{\rm tot} +  \sum_{i = n,u,d}\mu_i n_i\hspace{3mm}, \label{eq24}
\end{equation}
\begin{equation}
\frac{c_s}{c}=\sqrt{\frac{\partial P_{\rm tot}}{\partial {\cal E}_{\rm tot}}}
\label{speed-2}  
\end{equation}
%%%%%%%%%%%%%%%%%%%%%%%%%
The quantities $n_B$, ${\cal E}_{\rm tot}$, $P_{\rm tot}$ are functions of the momentum $k_{F_n}$. In totally, from Eqs.~(\ref{eq22})-(\ref{speed-2}),  the equation of state and the speed of sound  for the quarkyonic matter can be readily determined. 

\begin{figure}

\includegraphics[width=245pt,height=19pc]{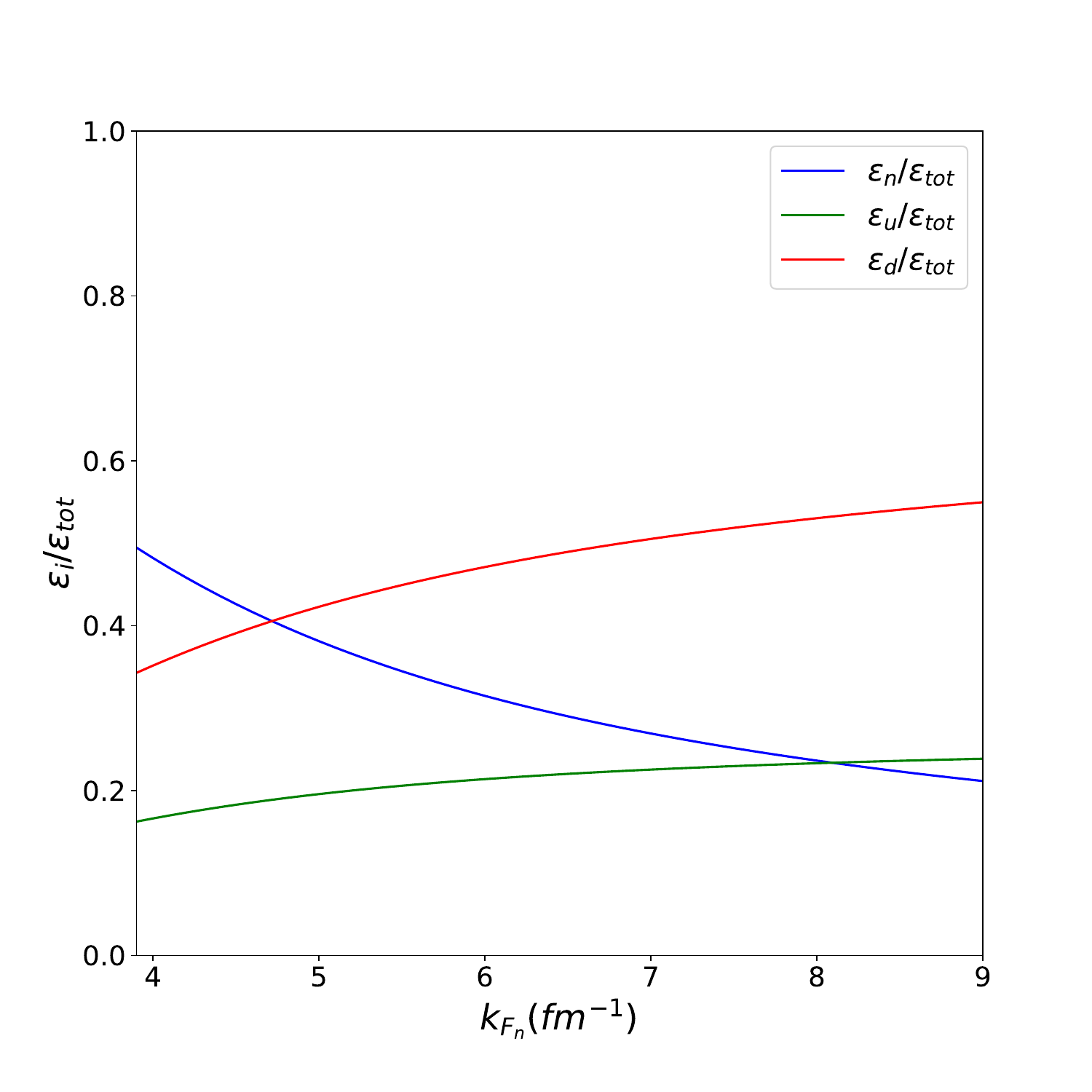}
\caption{The fraction of the contribution to the total energy density from the energy density of neutrons, up quarks, and down quarks as a function of the neutrons Fermi momentum ($k_{F_n}$) (blue, green and red lines respectively), for $n_{\rm tr}= 0.2$ fm$^{-3}$, $\Lambda_{Qyc} = 160$ MeV and $\kappa = 0.3$.}
\label{e_i-e_tot}
\end{figure}

\begin{figure}

\includegraphics[width=245pt,height=19pc]{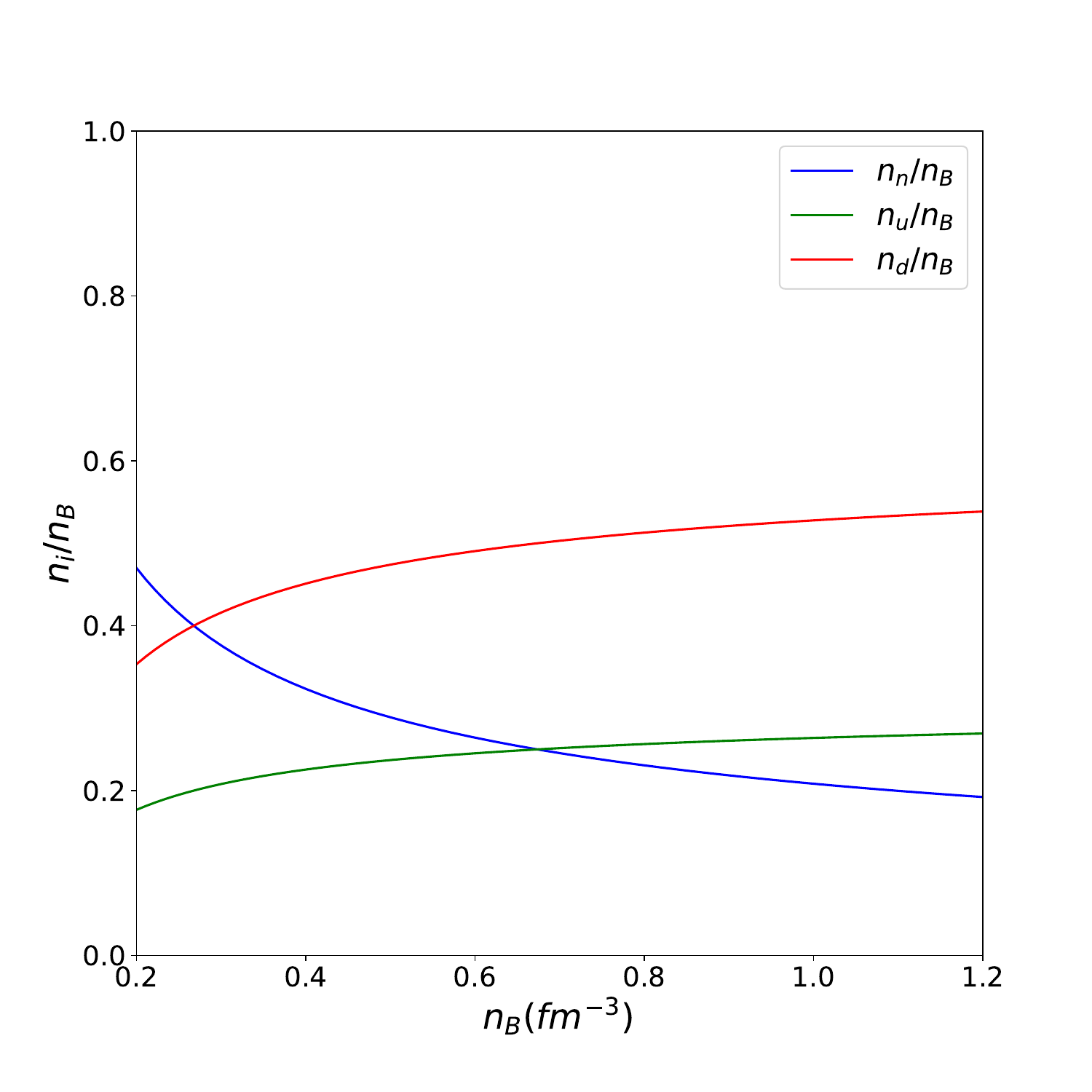}
\caption{The fractions of neutrons, up and down quarks as a function of the total baryon density ($n_B$) (blue, green and red lines respectively), for $n_{\rm tr} = 0.2$ fm$^{-3}$, $\Lambda_{Qyc} = 160$  MeV and $\kappa = 0.3$.}
\label{ni-n_tot}
\end{figure}

In Fig.~\Ref{e_i-e_tot} we display
the fraction of the contribution to the total energy density from the energy density of neutrons, up quarks, and down quarks as a function of the neutron's Fermi momentum ($k_{F_n}$) for a specific parametrization of the EOS. It is clear that for small values of $k_{F_n}$ the contribution of neutrons dominates, but at higher momenta, the quark's contribution becomes more significant. Similarly, using the same parametrization of the EOS, in Fig.~\Ref{ni-n_tot}, we present the fractions of neutrons, up quarks, and down quarks as a function of the total baryon density $n_B$.

%%%%%%%%%%%%%%%%%%%%%%%%%%%%%%%%%%
\section{Neutron star structure}
%%%%%%%%%%%%%%%%%%%%%%%%%%%%%%%%%%%
Having now constructed the equation of state of neutron star matter we can calculate their basic properties (mass, radius and tidal deformability) by solving the Tolman-Oppenheimer-Volkoff (TOV) equations that express the hydrostatic equilibrium~\cite{Shapiro:1983du,Haensel2007NeutronS1}. In particular, the mechanical equilibrium of the neutron star matter is determined by the system of two differential equations (TOV) equations and  the equation of state ${\cal E}={\cal E}(P)$ of the fluid.
This system reads
\begin{eqnarray}
\frac{dP(r)}{dr}&=&-\frac{G{\cal E}(r) M(r)}{c^2r^2}\left(1+\frac{P(r)}{{\cal E}(r)}\right) \nonumber \\
&\times&
 \left(1+\frac{4\pi P(r) r^3}{M(r)c^2}\right) \left(1-\frac{2GM(r)}{c^2r}\right)^{-1}
\label{TOV-1}
\end{eqnarray}
%%%%
\begin{equation}
\frac{dM(r)}{dr}=\frac{4\pi r^2}{c^2}{\cal E}(r).
\label{TOV-2}
\end{equation}
%%%%%%%%%%%%%%%
The solving of the coupled differential equations (\ref{TOV-1}) and (\ref{TOV-2}) for $P(r)$ and $M(r)$ requires their numerical integration from the origin ($r = 0$) to the point $r=R$ where the pressure becomes practically zero. At this point the radius and the mass of the neutron star are computed.

In recent years, valuable information has been obtained from observations of gravitational waves resulting from mergers of black hole–neutron star and neutron star–neutron star binary systems. These sources enable the measurement of various properties of neutron stars. During the inspiral phase of binary neutron star systems, tidal effects become detectable. Specifically, the tidal Love number 
$k_2$  describes the neutron star's response to the tidal field and depends on both the neutron star mass and the applied equation of state (EoS). The exact relationship describing these tidal effects is given below~\cite{PhysRevD.77.021502,Hinderer_2008}
\begin{equation}
Q_{ij}=-\frac{2}{3}k_2\frac{R^5}{G}E_{ij}\equiv- \lambda E_{ij},
\label{Love-1}
\end{equation}
where $\lambda$ is the tidal deformability. The tidal Love number $k_2$ is given by ~\cite{PhysRevD.77.021502,Hinderer_2008}
\begin{eqnarray}
k_2&=&\frac{8\beta^5}{5}\left(1-2\beta\right)^2\left[2-y_R+(y_R-1)2\beta \right]\nonumber\\
& \times&
\left[\frac{}{} 2\beta \left(6  -3y_R+3\beta (5y_R-8)\right) \right. \nonumber \\
&+& 4\beta^3 \left.  \left(13-11y_R+\beta(3y_R-2)+2\beta^2(1+y_R)\right)\frac{}{} \right.\nonumber \\
&+& \left. 3\left(1-2\beta \right)^2\left[2-y_R+2\beta(y_R-1)\right] {\rm ln}\left(1-2\beta\right)\right]^{-1}
\nonumber
\label{k2-def}
\end{eqnarray}
where $\beta=GM/Rc^2$ is the compactness of a neutron star. The parameter $y_R$ is determined by the following differential equation~\cite{PhysRevD.77.021502,Hinderer_2008}
\begin{equation}
r\frac{dy(r)}{dr}+y^2(r)+y(r)F(r)+r^2Q(r)=0 
\label{D-y-1}
\end{equation}
$F(r)$ and $Q(r)$ are functions of the energy density ${\cal E}(r)$, pressure $P(r)$, and mass $M(r)$ defined as
\begin{equation}
F(r)=\left[ 1- \frac{4\pi r^2 G}{c^4}\left({\cal E} (r)-P(r) \right)\right]\left(1-\frac{2M(r)G}{rc^2}  \right)^{-1}
\nonumber
\label{Fr-1}
\end{equation}
and
\begin{eqnarray}
r^2Q(r)&=&\frac{4\pi r^2 G}{c^4} \left[5{\cal E} (r)+9P(r)+\frac{{\cal E} (r)+P(r)}{\partial P(r)/\partial{\cal E} (r)}\right]
\nonumber\\
&\times&
\left(1-\frac{2M(r)G}{rc^2}  \right)^{-1}- 6\left(1-\frac{2M(r)G}{rc^2}  \right)^{-1} \nonumber \\
&-&\frac{4M^2(r)G^2}{r^2c^4}\left(1+\frac{4\pi r^3 P(r)}{M(r)c^2}   \right)^2 \nonumber \\
&\times&\left(1-\frac{2M(r)G}{rc^2}  \right)^{-2}.
\nonumber
\label{Qr-1}
\end{eqnarray}
Eq.(~\ref{D-y-1}) must be solved numerically and self consistently with the TOV equations under the following boundary conditions: $y(0)=2$, $P(0)=P_c$ ($P_{c}$ denotes the central pressure), and $M(0)=0$. The numerical integration provides the value of $y_R=y(R)$, which is a basic ingredient for $k_2$.

In addition, an important and well measured quantity by the gravitational wave detectors, which can be treated as a tool to impose constraints on the EoS, is the dimensionless tidal deformability $\Lambda$, defined as~\cite{PhysRevD.77.021502,Hinderer_2008} 
\begin{equation}
    \Lambda=\frac{2}{3}k_2 \left(\frac{c^2R}{GM}\right)^5=\frac{2}{3}k_2 (1.473)^{-5}\left( \frac{R}{{\rm Km}} \right)^5\left(\frac{M_{\odot}}{M}  \right)^5
\end{equation}
%%%
We notice that the tidal deformability  $\Lambda$ is sensitive to the neutron star radius, hence can provide information for the low density part of the EoS, which is related also to the structure and properties of finite nuclei.

%%%%%%%%%%%%%%%%%%%%%%%%%%%%%%%%%%%%
\section{Results and Discussion}
\begin{figure}
\includegraphics[width=245pt,height=19pc]{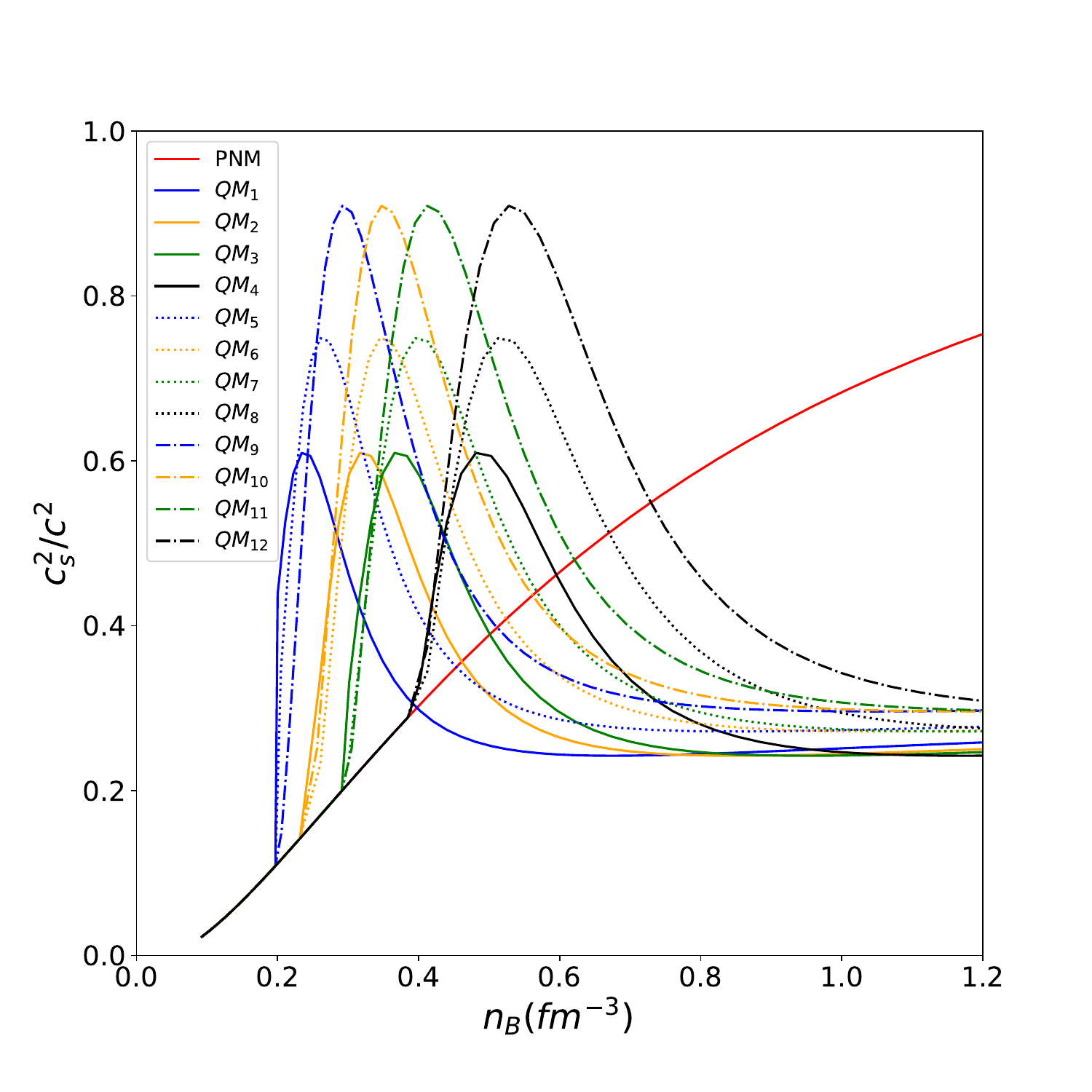}
\caption{The sound velocity for the quarkyonic model (QM)  for $n_{\rm tr} = 0.2,0.25,0.3,0.4$ fm$^{-3}$ (blue , yellow, green and black lines respectively) and for $\Lambda_{Qyc} = 160, 180, 200$ MeV (solid, dotted and dashed - dotted lines respectively). The solid red line corresponds to the pure neutron matter (PNM).}
\label{fig2}
\end{figure}

%%%%%%%%%%%%%%%%%%%%%%%%%%%%%%%%%%%%%%%%%
Firstly, we concentrate our study on predicting the speed of sound for various parametrizations of the equation of state, for the case of the NDU quarkyonic model (see Fig.~(\ref{fig2})). In particular, we employ several values of the transition density $n_{tr}$ as well as for the parameter $\Lambda_{Qyc}$. We set the parameter $\Lambda_{Qyc}$ to be 160, 180 and 200 MeV and for each value of  $\Lambda_{Qyc}$ we set the transition density to be 0.2, 0.25, 0.3 and 0.4 and we call each equation of state $QM_1$ - $QM_{12}$ respectively. It is important to note that for the different values of transition density, the pick in the speed of sound as a function of the baryon density is not affected at all. However, the location of the peak, in each case strongly depends on the transition density. This behavior of sound speed is a typical characteristic of quarkyonic matter. On the other hand, we can see that as the parameter $\Lambda_{Qyc}$ increases, the maximum value of the speed of sound increases so we have a violation of causality for values $\Lambda_{Qyc}> 210 \hspace{1mm}$ MeV.

\begin{figure}
\includegraphics[width=245pt,height=19pc]{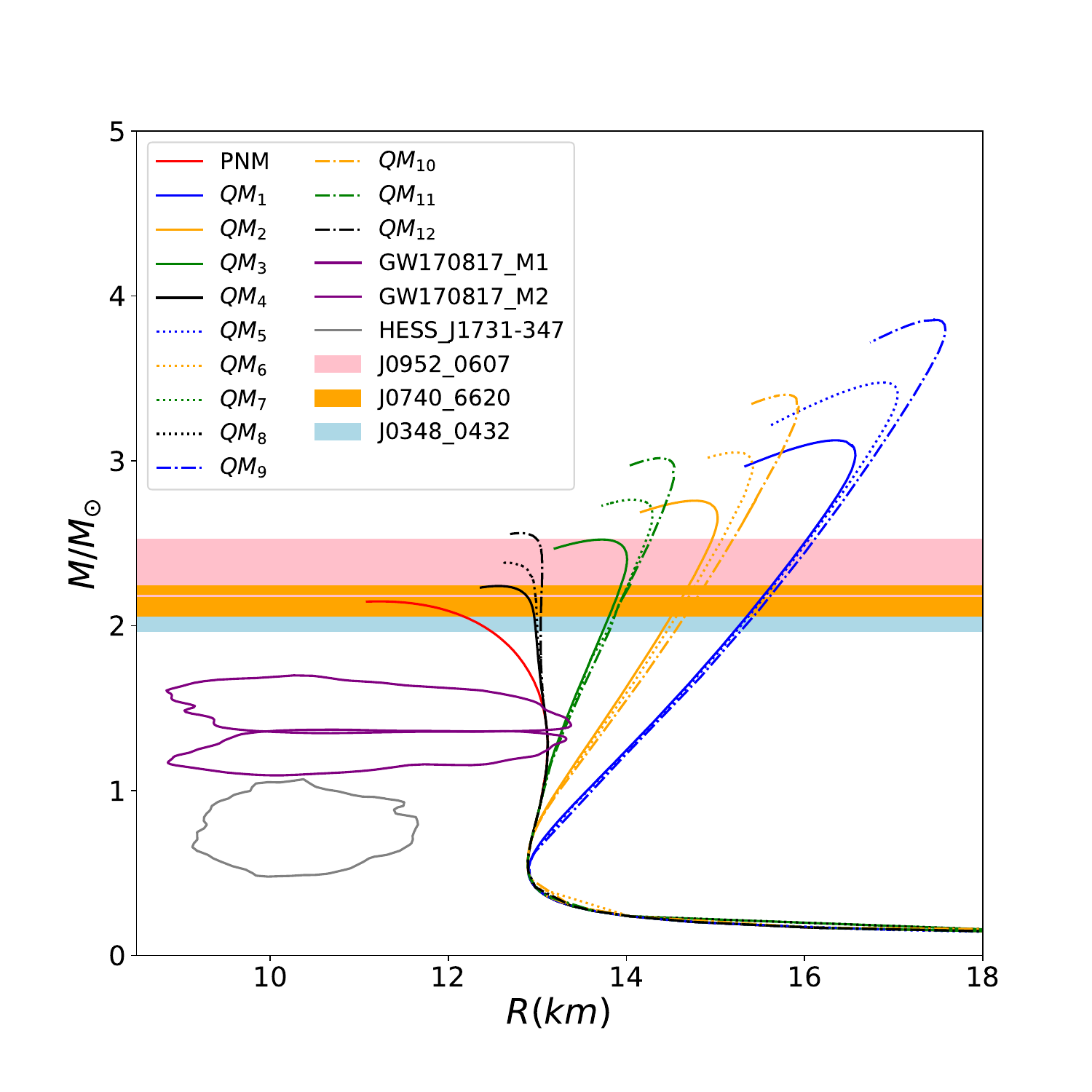}
\caption{M-R diagrams for the quarkyonic model (QM)  for $n_{\rm tr} = 0.2,0.25,0.3,0.4$ fm$^{-3}$ (blue, yellow, green and black lines respectively) and for $\Lambda_{Qyc} =160, 180, 200$  MeV (solid, dotted and dashed - dotted lines respectively). The solid red line corresponds to the pure neutron matter. The shaded regions correspond to possible constraints on
the maximum mass from the observation of PSR J0348+0432, PSR J0740+6620 and PSR J0952+0607 \cite{7712d035a7c043dd8e3205ecb6703f51,Antoniadis:2013pzd,PhysRevD.109.063017,2022ApJ...941..150S,Miller:2021qha}. The purple lines correspond to data resulting from LIGO and the grey one corresponds to data from the HESS observation \cite{2022NatAs...6.1444D}.}  
\label{mass-radius}
\end{figure}

The following step is to apply various equations of state to determine the fundamental properties of neutron stars and link them to the microscopic parameters of our model. We utilize the same set of equations of state that were previously used to predict the speed of sound. The results are displayed in Fig.~\ref{mass-radius}. Additionally, we incorporate recent observational data from the LIGO and HESS experiments to compare these findings with our own. As shown in Fig.~\ref{mass-radius}, the quarkyonic model for  $n_{\rm tr} =0.3$ and $0.4$  fm$^{-3}$
as well as the pure neutron matter model, predict that neutron stars with masses around $1.4$  solar masses have radii of approximately  ($13-13.3$) km. This prediction aligns well with the observational data estimates from LIGO. However, for lower values of  $n_{\rm tr}$ the predicted radii deviate from observational data, and the maximum masses significantly exceed observed values, leading to unrealistic results. Interestingly, for fixed values of $n_{\rm tr}$,
$\Lambda_{Qyc}$ parameter only influences the maximum mass, while the radius remains unaffected.

\begin{figure}

\includegraphics[width=245pt,height=19pc]{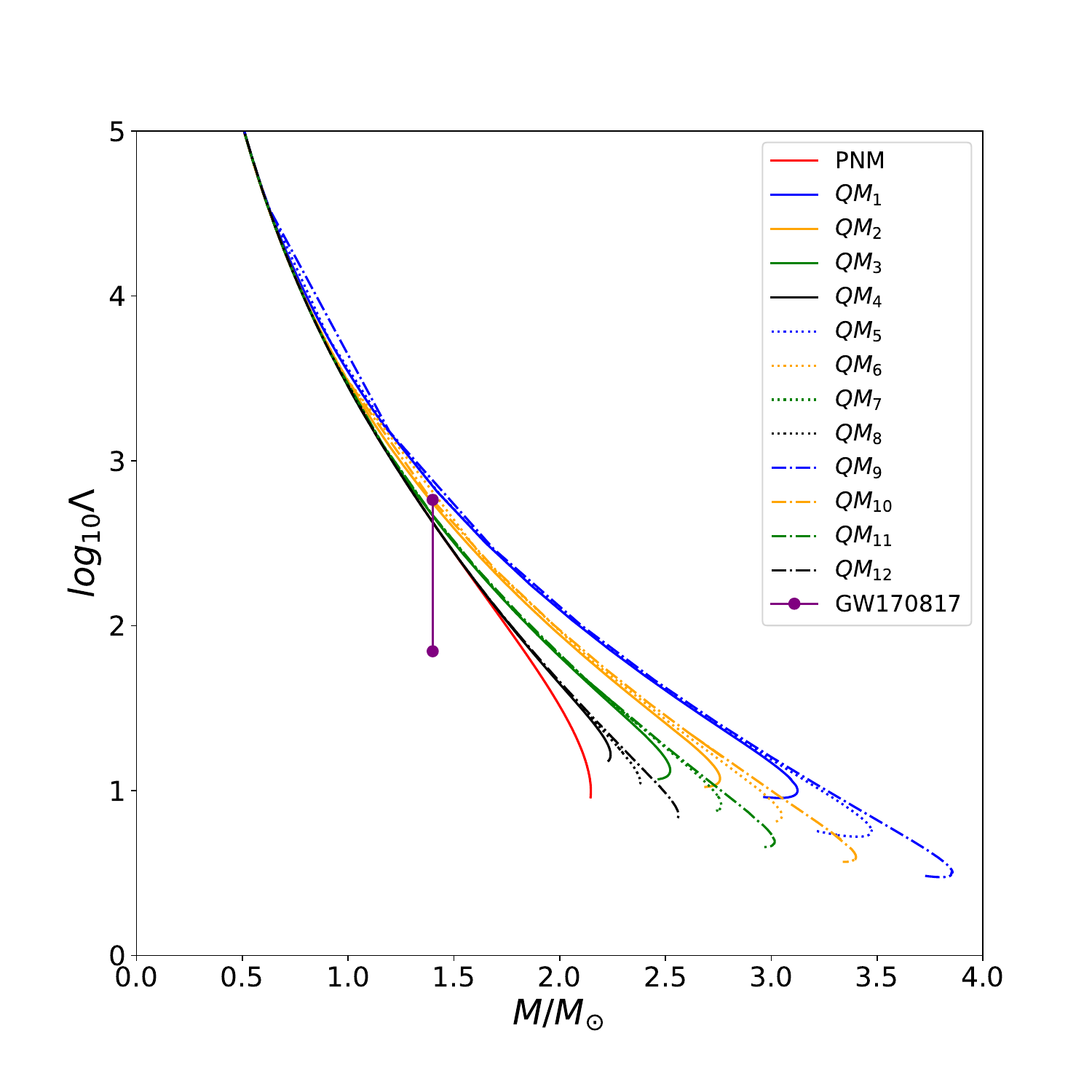}
\caption{Tidal deformability as a function of neutron star mass for the quarkyonic model (QM) for $n_{tr} = 0.2,0.25,0.3,0.4 \hspace{1mm}$fm$^{-3}$ (blue, yellow, green and black lines respectively) and for $\Lambda_{Qyc} = 160, 180, 200\hspace{1mm}$ MeV (solid, dotted and dashed - dotted lines respectively). The solid red line corresponds to the pure neutron matter model (PNM). The purple line corresponds to the event GW170817~\cite{Abbott_2017}.}
\label{tidal}
\end{figure}

\begin{figure}

\includegraphics[width=245pt,height=19pc]{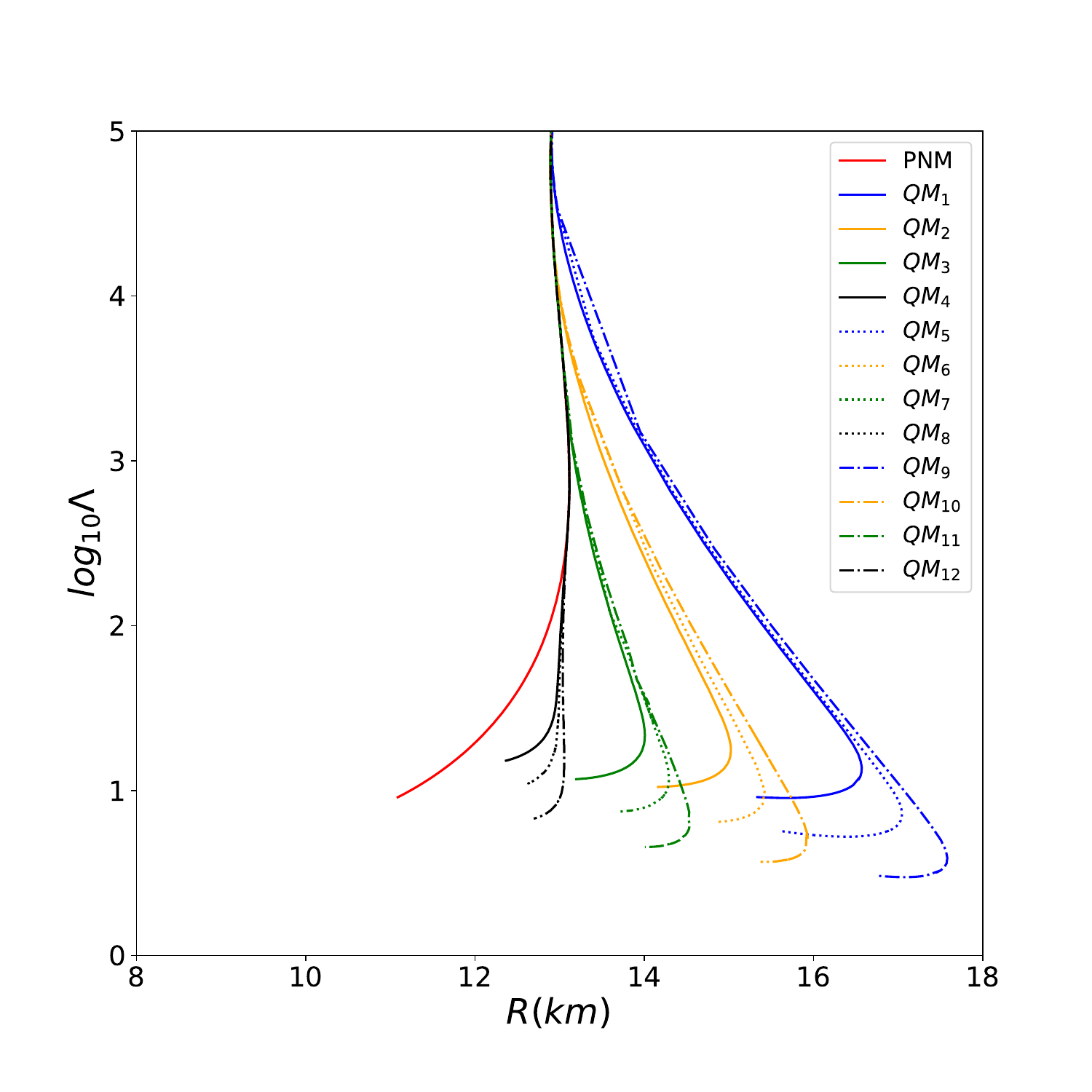}
\caption{Tidal deformability as a function of neutron star radius for the quarkyonic model (QM) for $n_{tr} = 0.2, 0.25, 0.3, 0.4$ fm$^{-3}$ (blue, yellow, green and black lines respectively) and for $\Lambda_{Qyc} = 160, 180, 200$ MeV (solid, dotted and dashed - dotted lines respectively). The solid red line corresponds to the pure neutron matter.}
\label{tidal vs radius}
\end{figure}

\begin{figure}
\includegraphics[width=245pt,height=19pc]{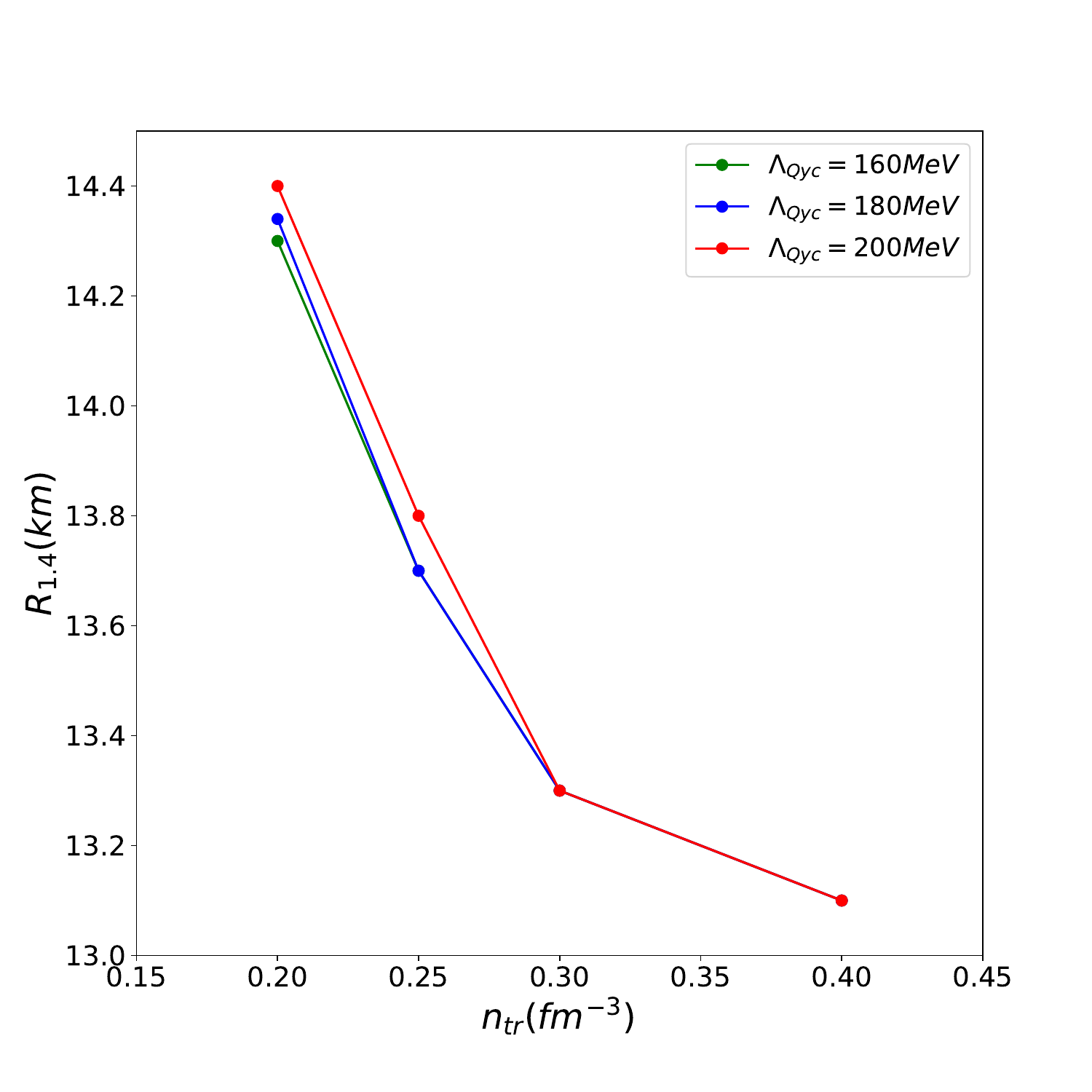}
\caption{The radius of a neutron star with mass $1.4M_{\odot}$ versus the transition density for $\Lambda_{Qyc}=160, 180, 200$ MeV (green, blue and red lines respectively).}
\label{R14 vs ntr}
\end{figure}

\begin{figure}
\includegraphics[width=245pt,height=19pc]{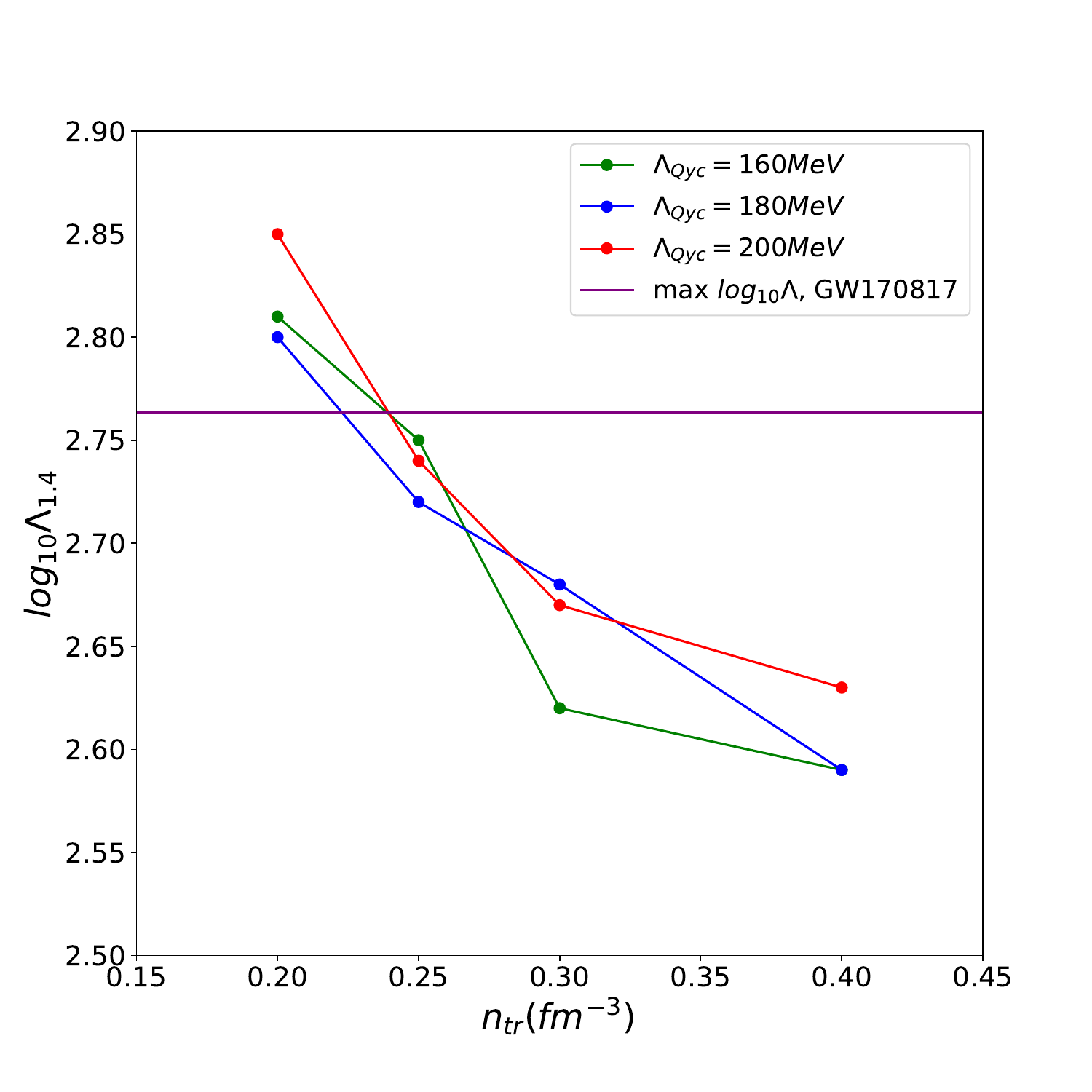}
\caption{The logarithm of tidal deformability for a neutron star with mass $1.4M_{\odot}$ versus the transition density for $\Lambda_{Qyc}=160, 180, 200$  MeV (green, blue and red lines respectively). The purple line corresponds to the maximum value of $\log_{10}\Lambda$ which arises from the event GW170817 (see Ref.~\cite{Abbott_2017}).}
\label{L14 vs ntr}
\end{figure}

Tidal deformability, a quantity primarily influenced by the radius (and to a lesser extent, the mass) of a neutron star, can be inferred from the measurement of gravitational waves emitted following a binary neutron star merger. This measurement is particularly useful for constraining the macroscopic properties of equations of state and for testing various nuclear models and approaches. In view of the above in Figs.~\ref{tidal} and \ref{tidal vs radius} we illustrate the logarithm of tidal deformability as a function of mass and the radius respectively for a neutron star, according to our models, for several values of the transition density and for $\Lambda_{Qyc} = 160, 180, 200$ MeV. 

For comparison, in  Fig.~\ref{tidal}, we include observational constraints from the GW170817 event (see Ref.~\cite{Abbott_2017}) for the tidal deformability corresponding to a neutron star with a mass of $1.4$ solar masses. Based on this, only the prediction for the case with a transition density of $n_{\rm tr}=0.2$ fm$^{-3}$ falls outside these constraints.

Finally, in Fig.~\ref{R14 vs ntr} and~\ref{L14 vs ntr} we summarize our results for the radius $R_{1.4}$ and tidal deformability $\Lambda_{1.4}$, correspond to a neutron star with mass $1.4$ solar masses, produced for the values of transition density and the parameter $\Lambda_{Qyc}$. An interesting feature is that the value of the parameter $\Lambda_{Qyc}$ does not affect the radius of a neutron star with this mass but results in greater maximum masses as it increases. On the other hand, it's important to note that the lower the transition density, the larger the radius of a $1.4 M_{\odot}$ neutron star. Based on the observational data, we can constrain the transition density to be greater than $0.3$ fm$^{-3}$.

\section{Concluding Remarks}
%%%%%%%%%%%%%%%%%%%%%%%%%%%%%%%%%%%%%%%
The main conclusions of the present study can be summarized as follows

\begin{enumerate}

\item For the first time, a quarkyonic model has been utilized that introduces interactions between nucleons in momentum space. This enhancement makes the quarkyonic model more comprehensive and flexible in describing quarkyonic matter and, by extension, the matter within neutron stars. Clearly, this interaction pertains exclusively to nucleons, while quarks are treated as a Fermi gas within the classical framework of the existing quarkyonic matter model.

\item It is worth noting that we have selected a model of nuclear matter with a fixed parameterization. While a more systematic study could be conducted by varying the specific parameterization, this was not the primary objective of this work. We concentrated on two fundamental parameters: the transition density  $n_{\rm tr}$ and the 
$\Lambda_{Qyc}$ parameter of quark matter. We investigated how these parameters affect the fundamental properties of neutron stars within the context of our specific model. Additionally, we made estimates regarding how these parameters might be determined through observational data.

\item To make the study more realistic, it should account for the fact that matter in neutron stars is in beta equilibrium, and therefore, at a minimum, protons and electrons should be included in the model. Such a study is currently underway. Moreover, we intend to further develop the model by incorporating temperature. This addition will enhance the model's realism, as studying neutron stars and specific processes such as mergers necessitates hot equations of state. We expect that future gravitational wave
observations from binary neutron star systems will give us information to constrain further some of the microscopic parameters of our model, so that to test and to
improve our equations of state.

\item Last but not least, it's important to note that quarkyonic matter may bridge the gap between hadronic and quark matter, potentially explaining a phase transition between these two states. While this theory is quite recent and many questions remain unanswered, it holds significant promise and offers numerous applications across various fields of theoretical physics.
In this direction, it is crucial to investigate whether there is any fundamental theory that can account for this state of matter~\cite{Cao:2022inx,PhysRevC.107.065201,Kovensky:2020xif,Kojo:2021ugu,2021arXiv210903887A}.

\end{enumerate}

\end{document}